\documentclass[12pt,onecolumn,preprint]{aastex6}
\usepackage{CJK}
\usepackage{color}

\usepackage{natbib}
\usepackage{hyperref}
\usepackage{cleveref}

\begin{document}
\title{\bf Bayesian Inference of the Symmetry Energy of Super-Dense Neutron-Rich Matter from Future Radius Measurements of Massive Neutron Stars}
\author{Wen-Jie Xie\altaffilmark{1} and Bao-An Li\altaffilmark{2}$^{*}$}
\altaffiltext{1}{Department of Physics, Yuncheng University, Yuncheng 044000, China}
\altaffiltext{2}{Department of Physics and Astronomy, Texas A$\&$M University-Commerce, Commerce, TX 75429, USA\\
\noindent{$^{*}$Corresponding author: Bao-An.Li@Tamuc.edu}}

\begin{abstract}
Using as references the posterior probability distribution functions (PDFs) of the Equation of State (EOS) parameters inferred from the radii of canonical neutron stars (NSs) reported by the LIGO/VIRGO and NICER Collaborations based on their observations of GW170817 and PSR J0030+0451, we investigate how future radius measurements of more massive NSs will improve our current knowledge about the EOS of super-dense neutron-rich nuclear matter, especially its symmetry energy term. Within the Bayesian statistical approach using an explicitly isospin-dependent parametric EOS for the core of NSs,
we infer the EOS parameters of super-dense neutron-rich nuclear matter from three sets of imagined mass-radius correlation data representing typical predictions by various nuclear many-body theories, i.e, the radius stays the same, decreases or increases with increasing NS mass within $\pm 15\%$ between 1.4 M$_{\odot}$ and 2.0 M$_{\odot}$. The corresponding NS average density increases quickly, slowly or slightly decreases as the NS mass increases from 1.4 M$_{\odot}$ to 2.0 M$_{\odot}$. While the EOS of symmetric nuclear matter (SNM) inferred from the three data sets are approximately the same, the corresponding symmetry energies above about twice the saturation density of nuclear matter are very different, indicating that the radii of massive NSs carry important information about the high-density behavior of nuclear symmetry energy with little influence from the remaining uncertainties of the SNM EOS at supra-saturation densities.
\end{abstract}
\maketitle
\newpage

\section{Introduction}

To understand the nature and constrain the Equation of State (EOS) of super-dense neutron-rich nuclear matter has been a major science goal shared by many astrophysical observations and terrestrial nuclear experiments, see, e.g., refs. \citep{Danielewicz02,LCK08,Lattimer16,Watts16,Oertel17,Ozel16,Li17,Herman17,Blaschke2018,Bom18,BUR18,ISAAC18,Pro19,Baiotti} for topical reviews.
The most basic quantity for calculating the EOS of nuclear matter at nucleon density $\rho=\rho_n+\rho_p$ and isospin asymmetry $\delta\equiv (\rho_n-\rho_p)/\rho$
is the average nucleon energy $E(\rho ,\delta )$
\begin{equation}\label{eos}
E(\rho ,\delta )=E_0(\rho)+E_{\rm{sym}}(\rho )\cdot \delta ^{2} +\mathcal{O}(\delta^4)
\end{equation}
according to essentially all existing nuclear many-body theories \citep{Bom91}. The first term $E_0(\rho)$ is the nucleon energy in symmetric nuclear matter (SNM) having equal numbers of neutrons and protons while the symmetry energy $E_{\rm{sym}}(\rho )$ quantifies the energy needed to make nuclear matter more neutron rich. While much progress has been made over the last few decades in constraining the SNM EOS in a broad density range, the symmetry energy $E_{\rm{sym}}(\rho )$ is relatively well constrained only around and below the saturation density of nuclear matter $\rho_0\approx 2.8\times 10^{14}$ g/cm$^{3}$ (0.16 fm$^{-3}$) \citep{Li98,ibook01,Bar05,Steiner05,LCK08,Lattimer2012,Tsang12,Dutra12,Dutra14,Chuck14,Tesym,Bal16,Li18}. While very little is known about the symmetry energy at supra-saturation densities. In fact, the $E_{\rm{sym}}(\rho )$ has been broadly recognized as the most uncertain part of the EOS of super-dense neutron-rich nucleonic matter, see, e.g., refs. \citep{Kut93,Ditoro2,BALI19}.

The nuclear symmetry energy has broad ramifications for many properties of neutron stars and gravitational waves from their mergers. For example, the density profile of isospin asymmetry in NSs at $\beta$ equilibrium, i.e, $\delta(\rho)$ or the corresponding proton fraction $x_p(\rho)$, is uniquely determined by the $E_{\rm{sym}}(\rho )$ through the $\beta$-equilibrium and charge neutrality conditions. Once the $\delta(\rho)$ is determined by the $E_{\rm{sym}}(\rho )$, both the pressure $P(\rho, \delta)$ and energy density $\epsilon(\rho,\delta)$ reduce to functions of nucleon density only. Their relation $P(\epsilon)$ can then be used to study NS structures.  Moreover, both the critical nucleon density $\rho_c$ (where $x_p(\rho_c)\approx 1/9$ in the neutron+proton+electron ($npe$) matter in NSs) above which the fast cooling of protoneutron stars by neutrino emissions through the direct URCA process can occur, and the crust-core transition density in NSs depend sensitively on the $E_{\rm{sym}}(\rho )$ \citep{Lattimer00}. Furthermore, the frequencies and damping times of various oscillations, especially the g-mode of the core and the torsional mode of the crust, quadrupole deformations of isolated NSs and the tidal deformability of NSs in inspiraling binaries also depend on the $E_{\rm{sym}}(\rho )$ \citep{DongLai,Plamen1,Newton,Wen19}. There is also a degeneracy between the EOS of super-dense neutron-rich matter and the strong-field gravity in understanding both properties of super-massive NSs and the minimum mass to form black holes. Thus, a precise determination of the $E_{\rm{sym}}(\rho )$ has broad impacts in many areas of astrophysics, cosmology and nuclear physics \citep{Wen09,XTHe}.

While it is very challenging to extract the density dependence of nuclear symmetry energy $E_{\rm{sym}}(\rho )$ from terrestrial experiments and/or astrophysical observations for many scientific and technical reasons, much progress has been made over the last two decades. For example, by 2013 there were at least 28 analyses of terrestrial nuclear laboratory experiments and astrophysical observations to extract the magnitude $E_{\rm{sym}}(\rho_0)$ and slope $L(\rho_0)$ of symmetry energy at saturation density $\rho_0$. Assuming all studies are equally reliable/respectable, these analyses together indicate that the fiducial values of $E_{\rm{sym}}(\rho_0)$ and $L(\rho_0)$ are, respectively, $E_{\rm{sym}}(\rho_0)=31.6\pm 2.66$ MeV and $L=59\pm 16$ MeV\,\citep{Li13}. In 2016, a survey of 53 analyses \citep{Oertel17} found the new fiducial values are $E_{\rm{sym}}(\rho_0)=31.7 \pm 3.2$~MeV and $L = 58.7 \pm 28.1$~MeV, respectively.
These results are consistent with the earlier ones albeit with a larger uncertainty for $L$ as more diverse analyses were included. Most of the post-GW170817 analyses of NS star radii and/or tidal deformability
found $L$ values are generally consistent with the above values, see, e.g., refs. \citep{BALI19,Baiotti} for reviews. Interestingly, predictions of some of the latest state-of-the-art microscopic nuclear many-body theories are in very good agreement with the above fiducial values. For example, using a novel Bayesian approach to quantify the truncation errors in chiral effective field theory (EFT) predictions for pure neutron matter and a many-body perturbation theory with consistent nucleon-nucleon and three-nucleon interactions up to fourth order in the EFT expansion, the $E_{\rm{sym}}(\rho_0)$ and $L(\rho_0)$ are found to be $E_{\rm{sym}}(\rho_0)=31.7 \pm 1.1$~MeV and $L = 59.8 \pm 4.1$~MeV, respectively \citep{Ohio20}. It thus seems that the values of $E_{\rm{sym}}(\rho_0)$ and slope $L(\rho_0)$ at saturation density $\rho_0$ are converging nicely while there are certainly needs to better understand and reduce both the statistical and systematic errors. It is worth noting that at sub-saturation densities, various studies on nuclear structures and reactions are also making significant progress \citep{Jorge14,Colo14,X18}. In particular, many studies of neutron-skins of heavy nuclei using various approaches including the parity violating electron-nucleus scatterings provide constrain on the symmetry energy around $2/3\rho_0$, see, e.g., refs. \citep{ZZ13,Vin14,India}. The latter has its own importance in both nuclear physics and astrophysics and can be extrapolated to somewhat higher densities near $\rho_0$.

\begin{figure}[!hpbt]
\centering
\vspace{1.cm}
\includegraphics[scale=0.5]{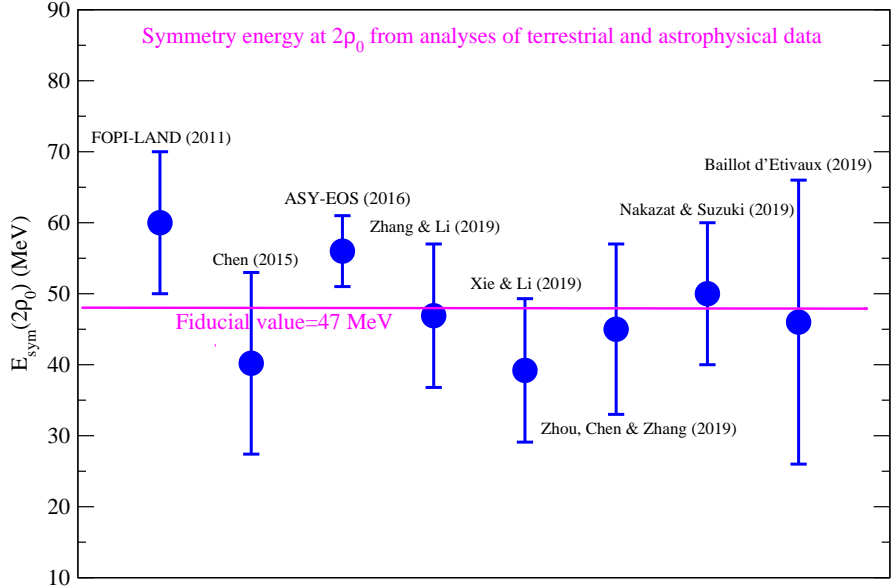}
\vspace{0.5cm}
\caption{Nuclear symmetry energy at twice the saturation density of nuclear matter deduced from energetic heavy-ion reactions in terrestrial laboratories and observations of neutron stars, see, text for details.}
\label{Esym-survey}
\end{figure}

It is also very encouraging to note that recent analyses of heavy-ion reaction experiments in terrestrial laboratories and properties of neutron stars from multiple messengers have led to some new progress in constraining the $E_{\rm{sym}}(\rho )$ up to about twice the saturation density.  For example, shown in Fig. \ref{Esym-survey} are the values of symmetry energy at $2\rho_0$, i.e.,  $E_{\rm{sym}}(2\rho_0)$, from (1) the FOPI-LAND \citep{Rus11} and (2) the ASY-EOS \citep{Rus16} Collaborations by analyzing the relative flows and yields of light mirror nuclei as well as neutrons and protons in heavy-ion collisions at beam energies of 400 MeV/nucleon, (3) (Chen) an extrapolation of the systematics of low-density symmetry energy \citep{ChenLW15}, (4) (Zhang \& Li) direct inversions of observed NS radii, tidal deformability and maximum mass in the high-density EOS space \citep{Zhang18,Zhang19epj,Zhang19apj}, (5) (Xie \& Li) a Bayesian inference from the radii of canonical NSs observed by using X-rays and gravitational waves from GW170817 \citep{Xie19}, (6) (Zhou, Chen \& Zhang) analyses of NS radii, tidal deformability and maximum mass within an extended Skyrme Hartree-Fock approach (eSHF) \citep{LWChen19,YZhou19}, (7) (Nakazato \& Suzuki) analyzing cooling timescales of protoneutron stars as well as the radius and tidal deformability of GW170817 \citep{Nakazato19}, (8) a Bayesian inference directly from the X-ray data of 7 quiescent low mass X-ray binaries in globular clusters \citep{Baillot19}. Despite of the rather different assumptions and methods used in analyzing the different types of laboratory and observational data,
it is very interesting to see that they all together are consistent with a fiducial value of $E_{\rm{sym}}(2\rho_0)=47$ MeV within the still relatively large error bars of the individual analyses.
Moreover, several recent theoretical studies also predicted values of $E_{\rm{sym}}(2\rho_0)$ consistent with its fiducial value of 47 MeV.  For example, an upper bound of  $E_{\rm{sym}}(2\rho_0) \leq 53.2$ MeV was derived recently in Ref. \citep{PKU-Meng} by studying the radii of neutron drops using the state-of-the-art nuclear energy density functional theories. Quantum Monte Carlo calculations using local interactions derived from chiral EFT up to next-to-next-to-leading order predicted a value of $E_{\rm{sym}}(2\rho_0) \approx 46 \pm 4$ MeV \citep{Diego}. While the latest many-body perturbation theory calculations with consistent nucleon-nucleon and three-nucleon interactions up to fourth order in the EFT expansion predicted a  value of $E_{\rm{sym}}(2\rho_0) \approx 45 \pm 3$ MeV \citep{Ohio20}.
They are both consistent with the fiducial value of $E_{\rm{sym}}(2\rho_0)=47$ MeV and have much smaller error bars. It is worth noting that the chiral EFT is currently applicable to a maximum density of about $2\rho_0$.

So, what is the main remaining problem with the symmetry energy of super-dense neutron-rich matter? Besides the large error bars of $E_{\rm{sym}}(2\rho_0)$ shown in Fig. \ref{Esym-survey}, detailed analyses
by both inverting directly the radii and/or tidal deformability of canonical NSs in the high-density EOS parameter space \citep{Zhang19epj,Zhang19jpg,Zhang19apj} or the Bayesian statistical inference of these NS observables \citep{Xie19} have shown clearly that the macroscopic properties of canonical NSs do not constrain the symmetry energy at densities above about $2\rho_0$. In particular, as we shall demonstrate, the skewness of symmetry energy characterizing its behavior above $2\rho_0$ is not constrained by the radii and/or tidal deformability of canonical NSs with masses around 1.4 M$_{\odot}$. This is mainly because
both the radii and tidal deformability of these NSs are mostly sensitive to the pressure at densities around $(1-2)\rho_0$ \citep{Lattimer00}. It was demonstrated clearly within both relativistic mean-field and Skyrme Hartree-Fock energy density functional theories that both the tidal deformability \citep{Fattoyev13} and radii \citep{Fattoyev14} of NSs heavier than 1.4M$_{\odot}$ have much stronger sensitivity to the high-density behavior of nuclear symmetry energy.

\begin{figure*}[htb]
\begin{center}
\resizebox{0.8\textwidth}{!}{
  \includegraphics{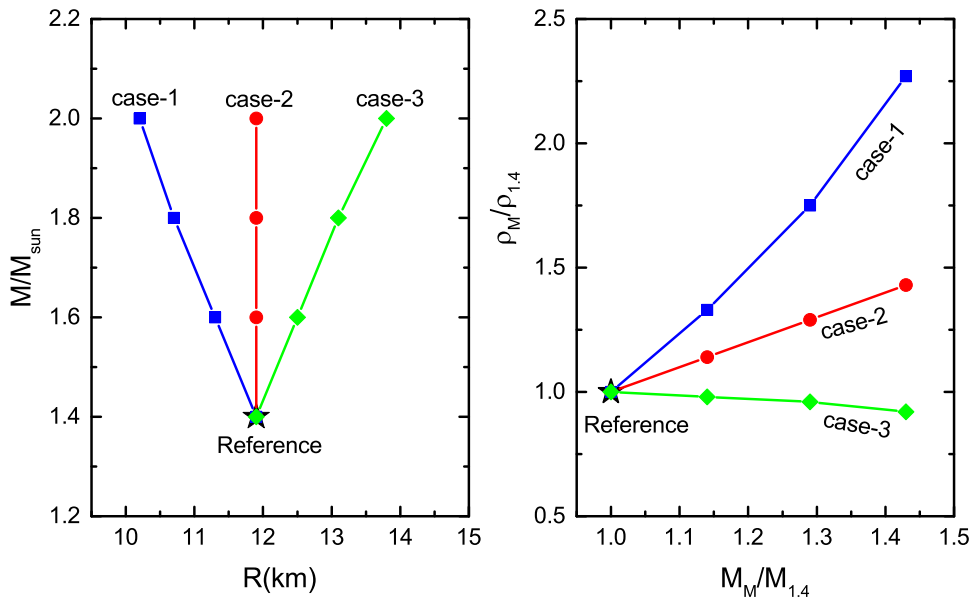}
  }
\caption{Left: representative mass-radius correlations considered for massive NSs. Right: the corresponding average density in NSs of mass M scaled by that of canonical NSs of mass M$_{1.4}\equiv$1.4M$_{\odot}$ as a function of the mass ratio $M/M_{1.4}$.}\label{MRD}
\end{center}
\end{figure*}

\begin{deluxetable}{lllll}
\tablecolumns{5}
\tablecaption{Imagined massive NS radii at 90\% confidence level}
\tablewidth{0pt}
\tablehead{&\colhead{$R_{1.4}$} & \colhead{$R_{1.6}$} & \colhead{$R_{1.8}$}& \colhead{$R_{2.0}$ (km)}
}
\startdata
\label{tab-data}
Reference & 11.9$\pm1.4$  & & & \\
case-1 & 11.9$\pm1.4$ &11.3$\pm1.4$ &10.7$\pm1.4$ &10.2$\pm1.4$  \\
case-2 & 11.9$\pm1.4$ &11.9$\pm1.4$ &11.9$\pm1.4$ &11.9$\pm1.4$  \\
case-3 & 11.9$\pm1.4$ &12.5$\pm1.4$ &13.1$\pm1.4$ &13.8$\pm1.4$  \\
\enddata
\end{deluxetable}

So, what is new in this work? It was speculated earlier that to constrain the symmetry energy significantly above $2\rho_0$,  one may have to study the radii of more massive NSs and/or additional messengers especially those directly from NS cores or emitted during collisions between either two NSs in space or two heavy nuclei in the laboratory \citep{BALI19,Xie19}. Using as references the posterior probability distribution functions (PDFs) of EOS parameters as well as the corresponding $E_{0}(\rho)$ and $E_{\rm{sym}}(\rho )$ determined by the GW170817 and recent NICER data for PSR J0030+0451 within a Bayesian statistical approach, here we examine how future radius measurements for massive NSs in the region of 1.4M$_{\odot}$ to 2.0M$_{\odot}$ may provide useful new information about the EOS especially its symmetry energy term at densities above $2\rho_0$.  More specifically, we use as imagined data in our Bayesian analyses the radii of three massive NSs of mass 1.6 M$_{\odot}$, 1.8 M$_{\odot}$ and 2.0 M$_{\odot}$ together with the reference radius $R_{1.4}$ for canonical NSs from GW170817 \citep{LIGO18} as listed in Table \ref{tab-data} along the three representative lines shown in the left window of Fig. \ref{MRD}. The radius as a function of mass along the three lines can be described approximately by
\begin{equation}\label{MRD-e}
R(M)~ ({\rm km})=\left
\{\begin{array}{ll}
R_{1.4}-4.2(M/M_{1.4}-1),~~&{\rm case-1,}\\
R_{1.4}=11.9\pm 1.4,~~&{\rm case-2,}\\
R_{1.4}+4.2(M/M_{1.4}-1),~~&{\rm case-3.}
\end{array}\right.
\end{equation}
The corresponding average densities scaled by that of a canonical NS of mass 1.4 M$_{\odot}$, i.e.,
\begin{equation}
\rho_{M}/\rho_{1.4}\equiv (M/M_{1.4})\cdot (R_{1.4}/R_M)^3,
\end{equation}
are shown in the right window for the three cases considered. In the case-1 where the radius decreases within increasing masses as predicted by many models, the average density increases by a factor of more than 2 going from canonical to 2 M$_{\odot}$ NSs, providing the best chance of probing super-dense NS matter. In the case-2, the radius is independent of mass and the average density increases with increasing mass relatively slowly completely due to the increase in mass. This case is also predicted by many theories and this assumption was actually used in a number of analyses of X-ray data. The case-3 is often predicted by models considering strangeness and/or hadron-quark phase transitions in NSs. In this case, the average density decreases slightly with increasing NS mass. All together, the three cases represent diverse model predictions. Moreover, the corresponding average density in NSs changes from case to case in a broad range. Of course, not all available theoretical predictions go through the reference point for canonical NSs as we require here.

While the latest NS maximum mass $M=2.14^{+0.10}_{-0.09}$~M$_\odot$ from observations of PSR~J0740+6620 \citep{M217} is rather precise, the available radius data of some massive NSs studied so far suffer from some systematic uncertainties, see, e.g., discussions in ref. \citep{Steiner18}. Fortunately, NICER and several more advanced X-ray observatories proposed are expected to measure much more precisely the radii of NSs in a broad mass range \citep{wp1,Strobe,wp2,Watts19}. It is thus useful to know what new physics can be extracted from future radius data of massive NSs compared to what we have already learned from studying the radii of canonical NSs. Moreover, if one considers the case-2 as the mean, the case-1 and case-3 as the lower and upper $1\sigma$ systematic error bounds of radius measurements, our imaginary data in Table \ref{tab-data} represent an approximately $\pm 15\%$ systematic error on top of the $\pm 4\%$ statistical error for NSs with mass 2.0 M$_{\odot}$. Comparing results of Bayesian inferences using the three typical cases will help us understand how the systematic errors in measuring the radii of massive NSs may affect what EOS information one can infer reliably.

So, what are the most important and interesting findings in this work? We find that the 68\% confidence boundaries of the SNM EOS $E_{0}(\rho)$ from the case-1 to case-2 and then the case-3 becomes only slightly more stiff, indicating that the $\pm 15\%$ systematic error in measuring the radii of massive NSs will not affect much the accuracy of extracting the SNM EOS at supra-saturation densities. While the corresponding symmetry energy $E_{\rm{sym}}(\rho )$ becomes significantly more stiff gradually. In particular, the PDFs of parameters characterizing the high-density $E_{\rm{sym}}(\rho )$ are significantly different, indicating that the radii of massive NSs have the strong potential of constraining tightly the $E_{\rm{sym}}(\rho )$ above $2\rho_0$ with little influence from the remaining uncertainties of SNM EOS at supra-saturation densities.

In the following, we shall first summarize the main ingredients of our Bayesian approach using an explicitly isospin-dependent parametric EOS for NSs containing neutrons, protons, electrons and muons (i.e., the $npe\mu$ model). We then establish the reference PDFs of EOS model parameters for canonical NSs using the radius data from LIGO/VIRGO and NICER Collaborations. We then compare the posterior PDFs and correlations of EOS parameters as well as the resulting 68\% confidence boundaries of the $E_{0}(\rho)$ and $E_{\rm{sym}}(\rho )$ for the three cases with respect to the reference. We also examine effects of the prior ranges of the poorly known high-density EOS parameters on inferring their posterior PDFs from the radii of massive NSs. A summary will be given at the end.

\section{Theoretical framework}
For completeness and ease of discussions, here we briefly recall the Bayesian inference approach using an isospin-dependent parametric EOS for the core of NSs consisting of nucleons, electrons and muons. More details can be found in our earlier publication \citep{Xie19}. The $npe\mu$ model is the minimum model for the core of NSs. Certainly, in super-dense matter new particles and/or phases may appear. Results of our study within the minimum model thus have to be understood within the model limitations. Nevertheless, we feel that our results establish a useful baseline for future studies including more degrees of freedom and new phases.

\subsection{Isospin-dependent parameterizations for the core EOS of NSs}
Within the $npe\mu$ model, the pressure is written in terms of the nucleon number density $\rho$ and isospin asymmetry $\delta$ as
\begin{equation}\label{pressure}
  P(\rho, \delta)=\rho^2\frac{d\epsilon(\rho,\delta)/\rho}{d\rho},
\end{equation}
where $\epsilon(\rho, \delta)=\epsilon_n(\rho, \delta)+\epsilon_l(\rho, \delta)$ denotes the energy density with $\epsilon_n(\rho, \delta)$ and $\epsilon_l(\rho, \delta)$ being respectively the energy densities of nucleons and leptons. While the $\epsilon_l(\rho, \delta)$ is calculated using the noninteracting Fermi gas model \citep{Oppenheimer39}, the $\epsilon_n(\rho, \delta)$ is related to the energy per nucleon $E(\rho, \delta)$ and the average mass of nucleons $M_N$ via
\begin{equation}\label{lepton-density}
  \epsilon_n(\rho, \delta)=\rho [E(\rho,\delta)+M_N].
\end{equation}
We parameterize the two parts of $E(\rho, \delta)$ according to
\begin{eqnarray}\label{E0para}
  E_{0}(\rho)&=&E_0(\rho_0)+\frac{K_0}{2}(\frac{\rho-\rho_0}{3\rho_0})^2+\frac{J_0}{6}(\frac{\rho-\rho_0}{3\rho_0})^3,\\
  E_{\rm{sym}}(\rho)&=&E_{\rm{sym}}(\rho_0)+L(\frac{\rho-\rho_0}{3\rho_0})+\frac{K_{\rm{sym}}}{2}(\frac{\rho-\rho_0}{3\rho_0})^2
  +\frac{J_{\rm{sym}}}{6}(\frac{\rho-\rho_0}{3\rho_0})^3\label{Esympara}
\end{eqnarray}
where $E_0(\rho_0)=-15.9 \pm 0.4$ MeV \citep{Brown14} is the nuclear binding energy at $\rho_0$.
As discussed in detail in refs. \citep{Zhang18,Zhang19epj}, these parameterizations are purposely chosen to have the same forms as if we are Taylor expanding known energy functionals. But they are just parameterizations of unknown functions. The parameters will be inferred (backward modeling) from Bayesian analyses of observational data, while in Taylor expansions they are calculated from known functions.

Compared to some other parameterizations widely used in the literature, such as the piece-wise polytropes for the pressure as a function of density that is composition-blind, by first parameterizing separately the $E_{0}(\rho)$ and $E_{\rm{sym}}(\rho )$ then reconstructing the pressure as a function of density at $\beta$ equilibrium, although being more complicated we can explore self-consistently the composition of super-dense neutron-rich matter. Actually, this is absolutely necessary to extract information about the symmetry energy at high densities. Moreover, parameterizing the $E_{0}(\rho)$ and $E_{\rm{sym}}(\rho )$ in Taylor forms has the advantage that we can directly use existing predictions of nuclear many-body theories and/or indications of nuclear experiments in setting the prior ranges of the EOS parameters to be inferred from astrophysical observations.  Mathematically, the two parameterizations naturally become the Taylor expansions of the unknown functions $E_{0}(\rho)$ and $E_{\rm{sym}}(\rho )$ when the density approaches $\rho_0$. Therefore, one may consider the above two expressions as having the dual meanings of being Taylor expansions around $\rho_0$ on one hand, and on the other hand being purely parameterizations far from $\rho_0$. Near $\rho_0$, the EOS parameters then obtain their asymptotic meaning one normally gives to the coefficients of Taylor expansions of known functions. Namely, the $K_0$ parameter represents the incompressibility of SNM $K_0=9\rho_0^2[\partial^2 E_0(\rho)/\partial\rho^2]|_{\rho=\rho_0}$ and the $J_0$ parameter represents the skewness of SNM $J_0=27\rho_0^3[\partial^3 E_0(\rho)/\partial\rho^3]|_{\rho=\rho_0}$ at saturation density. While the four parameters involved in the $E_{\rm{sym}}(\rho)$ denote the magnitude $E_{\rm{sym}}(\rho_0)$, slope $L=3\rho_0[\partial E_{\rm{sym}}(\rho)/\partial\rho]|_{\rho=\rho_0}$, curvature  $K_{\rm{sym}}=9\rho_0^2[\partial^2 E_{\rm{sym}}(\rho)/\partial\rho^2]|_{\rho=\rho_0}$ and skewness $J_{\rm{sym}}=27\rho_0^3[\partial^3 E_{\rm{sym}}(\rho)/\partial\rho^3]|_{\rho=\rho_0}$ of nuclear symmetry energy at saturation density, respectively. While in the literature, one usually uses the above asymptotic meanings to describe the EOS parameters, we emphasize again that they are parameters to be extracted from data through the Bayesian analyses. As such, the two parameterizations can be used far above $\rho_0$ as they are not simply Taylor expansions near $\rho_0$. Besides the obvious limitations on the flexibility and computing costs of using different number of parameters, Bayesian analyses also depend on the amount of relevant data available. We shall thus also investigate how using different numbers of parameters may affect what we extract from the three data sets by turning on and off the $J_{\rm{sym}}$ term in our Bayesian analyses.

We also note here that the density profile of isospin asymmetry $\delta(\rho)$ (or the corresponding proton fraction $x_p(\rho)$) at density $\rho$ within broad ranges of the symmetry energy parameters have been studied in detail in refs. \citep{Zhang18,Zhang19epj}. The relative particle fractions in NSs at $\beta$ equilibrium are obtained through the condition
$
  \mu_n-\mu_p=\mu_e=\mu_\mu
$
and the charge neutrality condition
$  \rho_p=\rho_e+\rho_\mu
$
for the proton density $\rho_p$, electron density $\rho_e$, and muon density $\rho_{\mu}$, respectively.
The chemical potential of particle $i$ is given by
$
  \mu_i=\frac{\partial\epsilon(\rho,\delta)}{\partial\rho_i}.
$
The most important information for this study is that soft/stiff symmetry energy at a given density will make the matter there more/less neutron-rich due to the $E_{\rm{sym}}(\rho )\cdot \delta ^{2}$ term in the EOS of isospin asymmetric matter in Eq. (\ref{eos}).

It is also worth noting that the parameterized $E_{0}(\rho)$ and $E_{\rm{sym}}(\rho )$ may not always go to zero mathematically as $\rho\rightarrow 0$ when the parameters are randomly selected in Bayesian analyses. Nevertheless, this does not create a physical problem as the parameterizations are only used in constructing the EOS for the core of NSs. In fact, we use the NV EOS \citep{Negele73} for the inner crust and the BPS EoS \citep{Baym71b} for the outer crust. The crust-core transition density and pressure are determined consistently from the same parametric EOS for the core. This is achieved by investigating the thermodynamical instability of the uniform matter in the NS core as detailed in ref. \citep{Zhang18}. Namely, when the incompressibility of the $npe\mu$ matter in the core becomes negative at low densities, the uniform matter becomes unstable against the formation of clusters, indicating the transition from core to crust \citep{Kubis04,Kubis07,Lattimer07,Xu09}.

\subsection{Bayesian inference}
As discussed above, the six parameters in Eqs. (\ref{E0para}) and (\ref{Esympara}) will be inferred from NS properties through Bayesian analyses. For completeness, we recall here the Bayesian theorem
\begin{equation}\label{Bay1}
P({\cal M}|D) = \frac{P(D|{\cal M}) P({\cal M})}{\int P(D|{\cal M}) P({\cal M})d\cal M},
\end{equation}
where the denominator is the normalization constant. The $P({\cal M}|D)$ represents the posterior PDF of the model $\cal M$ given the data set $D$. The $P(D|{\cal M})$ is the likelihood function obtained by comparing predictions of the model $\cal M$ with the data $D$, while the $P({\cal M})$ is the prior PDF of the model $\cal M$.
\begin{deluxetable}{lcc}
\tablecolumns{3}
\tablecaption{Prior ranges of the six EOS parameters used}\label{tab-pri}
\tablewidth{0pt}
\tablehead{
\colhead{Parameters (MeV)} & \colhead{Lower limit} & \colhead{Upper limit}
}
\startdata
$K_0$ & 220 & 260 \\
$J_0$ & -800 & 400 \\
$K_{\mathrm{sym}}$  & -400 & 100 \\
$J_{\mathrm{sym}}$ & -200 & 800 \\
$L$ & 30 & 90 \\
$E_{\mathrm{sym}}(\rho_0)$ & 28.5 & 34.9 \\
\enddata
\end{deluxetable}
The six EOS parameters are randomly sampled using flat prior PDFs between their minimum and maximum values listed in Table 2. The listed ranges are based on available indications of nuclear laboratory experiments and theoretical predictions. In particular,  the values of $K_0$, $E_{\rm sym}(\rho_0)$ and $L$ are known to be around $K_0\approx 240 \pm 20$ MeV \citep{Shlomo06,Piekarewicz10,Garg18}, $E_{\rm sym}(\rho_0)=31.7\pm 3.2$ MeV and $L\approx 58.7\pm 28.1 $ MeV \citep{Li13,Oertel17}, respectively. While the three high-density EOS parameters $K_{\rm{sym}}$, $J_{\rm{sym}}$ and $J_0$ are still poorly known to be around $-400 \leq K_{\rm{sym}} \leq 100$ MeV, $-200 \leq J_{\rm{sym}}\leq 800$ MeV, and $-800 \leq J_{0}\leq 400$ MeV \citep{Tews17,Zhang17}, respectively.

After generating the EOS parameters, $p_{i=1,2\cdots 6}$, one can construct the corresponding NS EOS model $\cal M$ as described earlier. Each NS EOS in the form of $P(\epsilon)$ is then used as an input to solve the Tolman-Oppenheimer-Volkov (TOV) NS structure equations \citep{Tolman34,Oppenheimer39}. The resulting mass-radius relation is then used in evaluating the likelihood of this set of EOS parameters. The radius data $D$ we shall use are summarized in Table \ref{tab-data}. The likelihood function measures the ability of the model $\cal M$ to reproduce the observational data. In the present work, we use
\begin{equation}\label{Likelihood}
  P[D|{\cal M}(p_{1,2,\cdots 6})]=P_{\rm{filter}} \times P_{\rm{mass,max}} \times P_{\rm{radius}},
\end{equation}
where the $P_{\rm{filter}}$ is a filter selecting EOS parameter sets satisfying the following conditions: (i) The crust-core transition pressure stays positive; (ii) At all densities, the thermaldynamical stability condition (i.e., $dP/d\varepsilon\geq0$) and the causality condition (i.e, the speed of sound is always less than that of light) are satisfied. The $P_{\rm{mass,max}}$ stands for the requirement that each accepted EOS has to be stiff enough to support the observed NS maximum mass $M_{\rm{max}}$. While in our previous work \citep{Xie19}, we have studied effects of using 1.97 M$_{\odot}$, 2.01 M$_{\odot}$ and 2.17 M$_{\odot}$ for $M_{\rm{max}}$ on extracting the EOS parameters, to be consistent with the reference data point $R_{1.4}=11.9\pm 1.4$ km extracted by the LIGO/VIRGO Collaborations from GW170817 by assuming $M_{\rm{max}}$=1.97 M$_{\odot}$ \citep{LIGO18}, we adopt the later in the present analysis. While the $P_{radius}$ is the probability for the chosen EOS model to reproduce the NS radius data. Depending on the number of data points in each data set along a mass-radius sequence (i.e, each case listed in Table \ref{tab-data}), the $P_{radius}$ may be a product of several Gaussian functions. It can be generally written as
\begin{equation}\label{Likelihood-radius}
  P_{radius}=\prod_{j=1}^{n}\frac{1}{\sqrt{2\pi}\sigma_{\mathrm{obs},j}}\exp[-\frac{(R_{\mathrm{th},j}-R_{\mathrm{obs},j})^{2}}{2\sigma_{\mathrm{obs},j}^{2}}],
\end{equation}
where $\sigma_{\mathrm{obs},j}$ represents the $1\sigma$ error bar of the observation $j$, and $n$ is the total number of data points used. For example, n is 1 for our reference where there is only one data point and 4 for the three cases listed in Table \ref{tab-data}.

A Markov-Chain Monte Carlo (MCMC) approach with the Metropolis-Hastings algorithm is used to simulate the posterior PDF of the model parameters. The PDFs of all individual EOS parameters and the two-parameter correlations are calculated by integrating over all other parameters. For example, the PDF for the $i$th parameter $p_i$ is given by
\begin{equation}\label{Bay3}
P(p_i|D) = \frac{\int P(D|{\cal M}) dp_1dp_2\cdots dp_{i-1}dp_{i+1}\cdots dp_6}{\int P(D|{\cal M}) P({\cal M})dp_1dp_2\cdots dp_6}.
\end{equation}
Numerically, we have to discard the initial samples in the so-called burn-in period because the MCMC does not sample from the equilibrium distribution in the beginning\citep{Trotta17}. It was found that 40,000 burn-in steps are enough as in our recent work\citep{Xie19}. We thus throw away the first 40,000 steps and use the remaining one million steps for calculating the posterior PDFs of the six EOS parameters in the present analysis.
\section{Results and discussions}
In this section, we present and discuss results of inferring the EOS parameters from the three sets of imagined radii of massive NSs. First we shall establish a reference using the radius data of canonical NSs measured by LIGO/VIRGO and NICER.

\begin{figure*}[htb]
\begin{center}
\resizebox{0.85\textwidth}{!}{
  \includegraphics{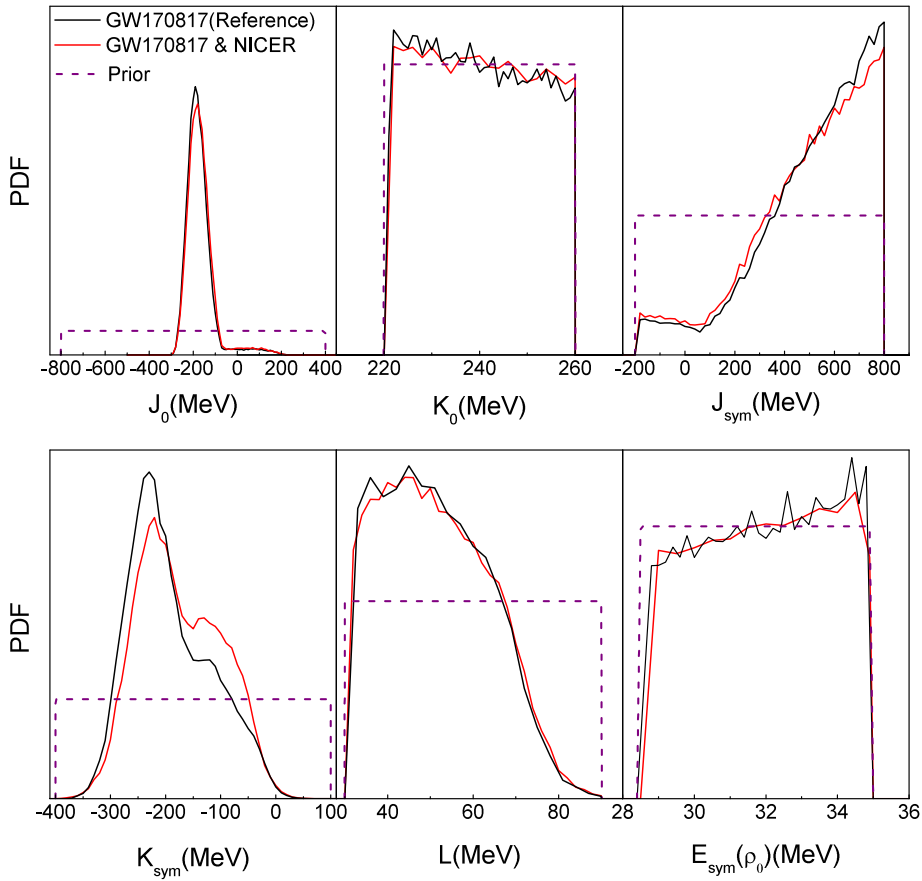}
  }
\caption{The posterior PDFs of NS EOS parameters from the two data sets indicated in comparison with their prior PDFs.}\label{nicer-fig1}
\end{center}
\end{figure*}

\subsection{Establish the reference: PDFs of EOS parameters from radii of canonical neutron stars}
After the historical observation of GW170817 binary NS merger event, an exciting flood of interesting papers appeared. Many studies using various approaches have extracted from the reported tidal deformability the radius $R_{1.4}$ of canonical NSs in the range of about 8.5 to 13.8 km, see, e.g., Fig. 41 in Ref. \citep{BALI19} for comparisons of the radii from 18 analyses carried out by mid-2019. The latest study combining multimessenger observations of GW170817 with many-body theory predictions using nuclear forces based on the chiral EFT found a more precise value of $R_{1.4}=11.0^{+0.9}_{-0.6}$ km at 90\% confidence level \citep{Capano20}. Most of the analyses indicate that the radius extracted is independent of the masses of the two NSs involved in GW170817. For example, the principal NS in GW170817 has a mass between 1.36 and 1.58 M$_{\odot}$, while the mass of the secondary NS  is between 1.18 and 1.36 M$_{\odot}$ \citep{LIGO18}.  Assuming initially the radii are mass dependent within two models in one of the first analyses \citep{LIGO18}, it was found that the radii of the two NSs are basically the same independent of their masses. After enforcing the requirement that all EOSs have to support NSs at least as massive as 1.97 M$_{\odot}$, both models lead to the same radius $R=11.9\pm 1.4$ km independent of the masses of the two NSs involved. In the independent analysis of GW170817 in ref. \citep{De18}, starting by explicitly assuming the two NSs have the same radius, the radius inferred was found independent of the prior Gaussian mass distributions centered around 1.33, 1.49 or 1.54 M$_{\odot}$. These studies clearly indicate that it is reasonable to assume that the radii of canonical NSs with masses around 1.4 M$_{\odot}$ are the same. Thus, as a reference for our study, we use $R_{1.4}=11.9\pm 1.4$ km from LIGO/VIRGO as a common and first data point as shown in Table 1.

It is exciting that the mass and radius of PSR J0030+0451 has been measured simultaneously recently by the NICER Collaboration. Two analyses of their data found the mass and radius are, respectively,
 $M=1.44^{+0.15}_{-0.14}$ M$_{\odot}$ and $R=13.02^{+1.24}_{-1.06}$ km \citep{Miller19}, and $M=1.34^{+0.16}_{-0.15}$ M$_{\odot}$ and $R=12.71^{+1.19}_{-1.14}$ km \citep{Riley19}.
 A recent study in ref. \citep{Zhang20} by directly inverting the radius in the high-density EOS parameter space of $J_0-K_{\rm{sym}}-J_{\rm{sym}}$ indicates that the NICER data provide similar constraints on the high-density EOS parameters as the NS tidal deformability from GW170817. As a comparison, we shall also calculate the PDFs of the six EOS parameters by combining the LIGO/VIRGO and NICER radius data. More specifically, for the NICER data we use the radius $R_{\mathrm{obs},j}=12.71$ km with $\sigma_{\mathrm{obs},j}$=1.16. In our model calculation with each EOS generated, we take the average radius for NSs with masses from 1.19 M$_{\odot}$ to 1.5 M$_{\odot}$ and regard it as the theoretical $R_{\mathrm{th},j}$ in evaluating the likelihood function using the NICER data.

Shown in Fig. \ref {nicer-fig1} are the posterior PDFs of the six EOS parameters using the GW170817 only and the combined GW170817+NICER data, respectively.
To see clearly the relative contributions of the likelihood and prior to the posterior PDFs, the uniform prior PDFs used in this work for the EOS parameters are also shown.
First of all, the two data sets lead to almost the same PDFs, indicating a strong consistency of the two observations. In the following discussions, we will use the results from using the GW170817 data alone as our reference. Secondly, the PDFs of $J_0$, $K_{\rm{sym}}$ and $L$ all have reasonably strong peaks. Compared to their flat prior PDF in the original ranges, obviously the radius data of canonical NSs have already constrained these parameters significantly with respect to their prior ranges.  However, the PDFs of the saturation-density parameters $K_0$ and $E_{\rm sym}(\rho_0)$ remain roughly the same as their prior PDFs. This is not surprising since the radius of canonical NSs are known to be most sensitive to the variation of pressure around $(1-2)\rho_0$ \citep{Lattimer00}. Perhaps, the most interestingly result is the PDF of the $J_{\rm{sym}}$ parameter which controls the behavior of nuclear symmetry energy above about $2\rho_0$. Overall, it favors a large positive value mostly because of its correlation with the $K_{\rm{sym}}$ which favors a large negative value. The shoulder in the PDF of $J_{\rm{sym}}$ in its negative region is due to its correlation with $J_0$ as we shall discuss in more detail later. Moreover, it is seen that the PDF of $J_{\rm{sym}}$ peaks at the upper end of its prior range, i.e,
$800$ MeV. We found that if we artificially enlarge its upper boundary, say to 1000 MeV, its most probable value will increase correspondingly to 1000 MeV. It indicates clearly that the radius data of canonical NSs do not constrain the $J_{\rm{sym}}$ parameter and the corresponding behavior of $E_{\rm sym}(\rho)$ above $2\rho_0$. This finding is consistent with the results from directly inverting the radius and/or the tidal deformability of canonical NSs in the $J_0-K_{\rm{sym}}-J_{\rm{sym}}$ high-density EOS space \citep{Zhang18,Zhang19epj,Zhang19apj}. This further illustrates the importance of investigating whether the radii of more massive NSs can do better.
\begin{figure*}[htb]
\begin{center}
\resizebox{0.8\textwidth}{!}{
  \includegraphics{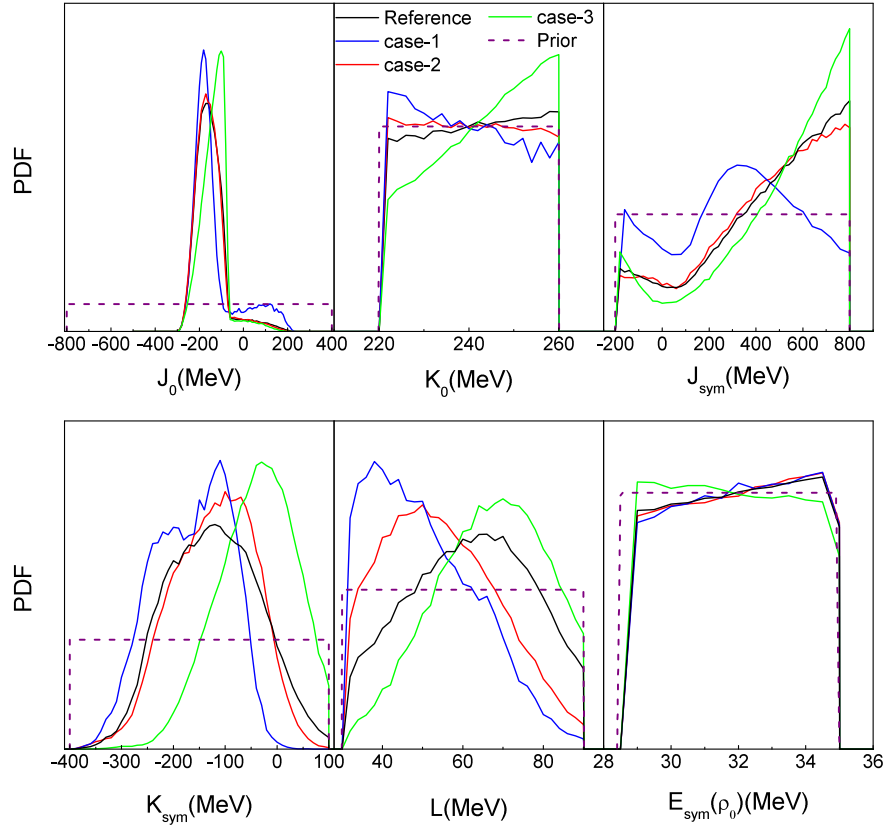}
  }
\caption{Posterior probability distribution functions of EOS parameters from the three sets of mass-dependent NS radius data shown in Fig. \ref{MRD} in comparison with their prior PDFs and the reference PDFs from GW170817 shown in Fig. \ref{nicer-fig1}.}\label{default-l1458}
\end{center}
\end{figure*}

\begin{figure*}[htb]
\begin{center}
\resizebox{0.497\textwidth}{!}{
  \includegraphics{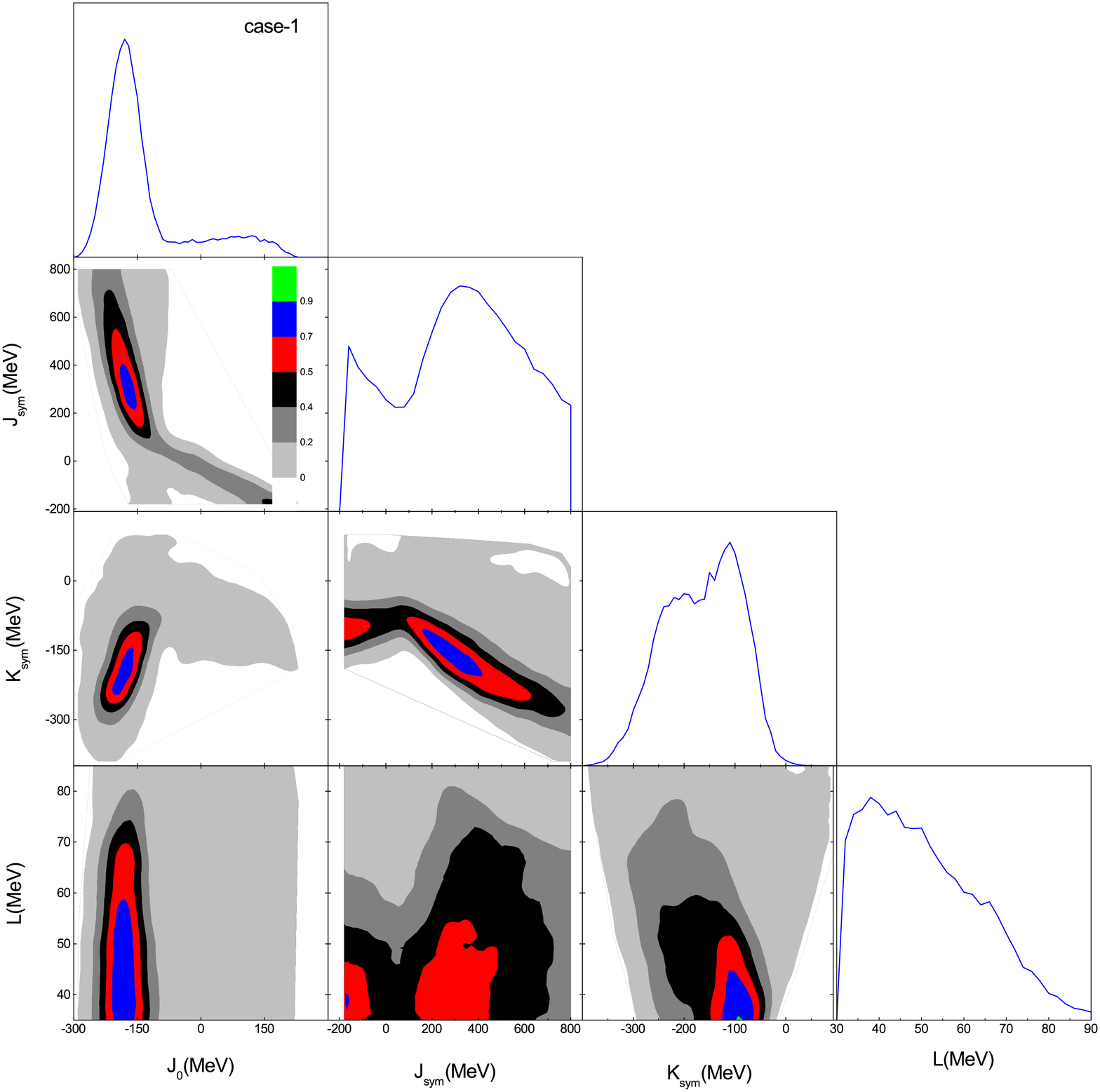}
  }
  \resizebox{0.497\textwidth}{!}{
  \includegraphics{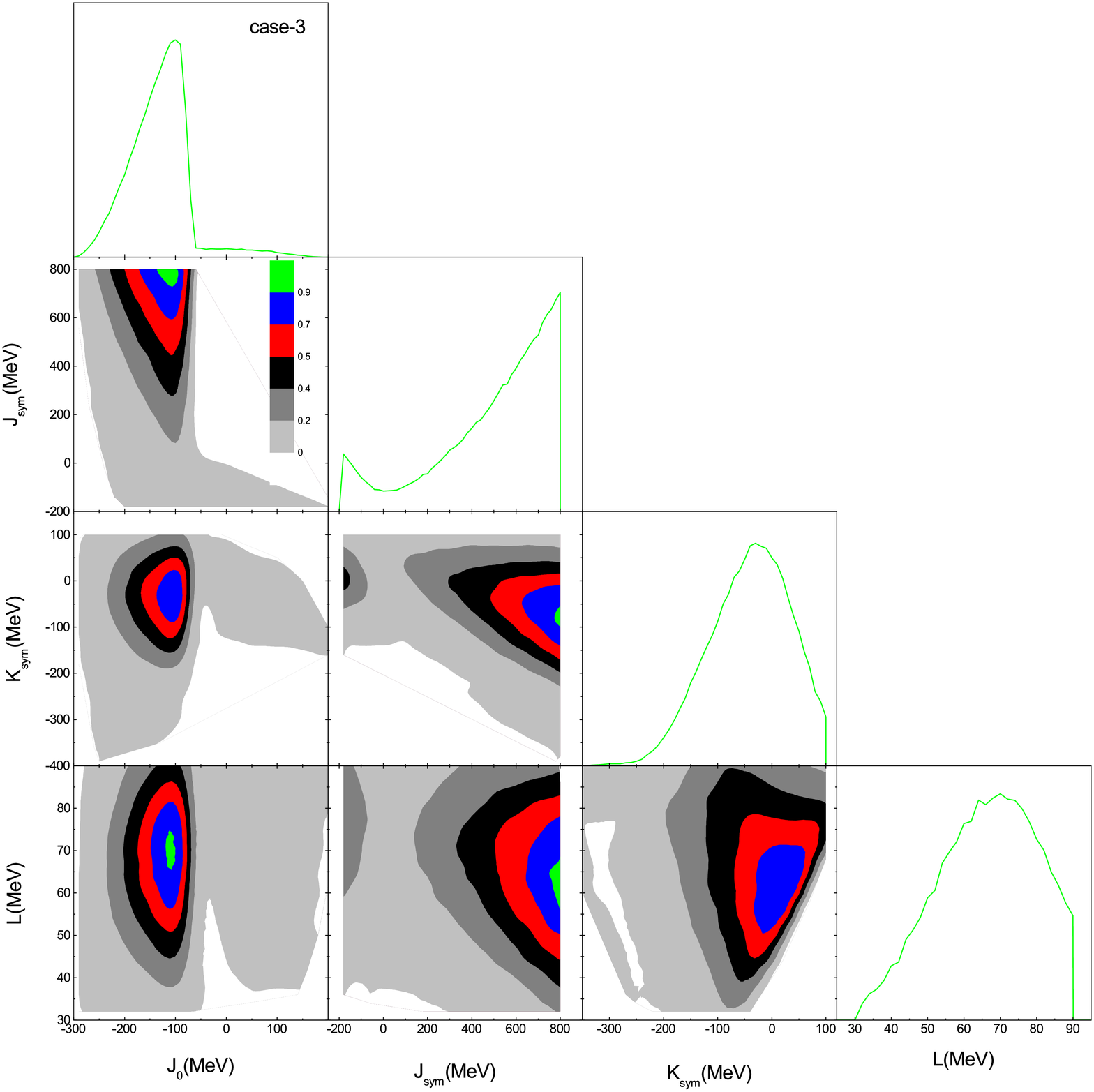}
  }
 \vspace{0.4cm}
\caption{Correlation functions of the high-density EOS parameters for the case-1 (left) and case-3 (right) described in the text.}\label{cor1}
\end{center}
\end{figure*}

\subsection{Posterior PDFs and correlations of EOS parameters from radii of massive neutron stars}
We now turn to inferring the PDFs of EOS parameters and their correlations from the imagined three radius data sets for massive NSs listed in Table 1.
Shown in Fig. \ref{default-l1458} are the posterior PDFs of the six EOS parameters derived from the three cases in comparison with the reference and their prior PDFs discussed in the previous subsection.

Firstly, the posterior PDFs of $E_{\rm sym}(\rho_0)$ in all cases remain approximately the same as its flat prior. This simply indicates that the radii of massive NSs are not sensitive to the value of symmetry energy at $\rho_0$ as one expects. For the case-2 where the radius is the same for all NSs considered, the PDFs of $J_0$, $J_{\mathrm{sym}}$ and $K_{\mathrm{sym}}$ are almost the same as for the reference from GW170817. While both $K_0$ and $L$ become smaller, indicating that both the SNM EOS $E_0(\rho)$ and symmetry energy $E_{\rm sym}(\rho)$ become slightly softer compared to the reference as we shall discuss in more detail. This observation is understandable. It was shown before that fixing the incompressibility $K_0$ of SNM but varying the slope $L$ of $E_{\rm sym}(\rho)$ at $\rho_0$ only changes the radii without changing the maximum mass of NSs, while fixing the symmetry energy but varying the $K_0$ only changes the NS maximum mass with little effect on the radii \citep{LiSteiner}. Already under the common constraint that all EOSs have to be stiff enough to support NSs at least as massive as 1.97 M$_{\odot}$, and as shown in Fig. \ref{MRD} in this case the average density only increases by at most 50\% going from NSs with 1.4 M$_{\odot}$ to 2.0 M$_{\odot}$, both  $K_0$ and $L$ only need to be slightly softened to support all the NSs with masses up to 2.0 M$_{\odot}$ but having the same radius. We emphasize that the pressure is required to always increase with increasing density. When the density is known to increase from a canonical NS to a heavier one, the required EOS can become softer as the pressure in a denser NS is naturally higher. All of these physical effects were incorporated consistently in the likelihood function discussed earlier. Therefore, it is understandable that the PDFs of $K_0$ and $L$ inferred from the mass-radius correlation of the case 2 shift towards lower values $K_0$ and $L$ slightly while the others remain approximately the same compared to the reference PDFs.

It is interesting to compare the results for case-1 and case-3.  The posterior PDFs of both the SNM EOS parameters $K_0$ and $J_0$ and the symmetry energy parameters $J_{\mathrm{sym}}$, $K_{\mathrm{sym}}$  and $L$ shift to the left (right) for case-1 (case-3), indicating that the SNM EOS $E_0(\rho)$ and symmetry energy $E_{\rm sym}(\rho)$ are softer (stiffer) for case-1 (case-3).
This can be understood from the relative densities reached in the two cases shown in the right window of Fig. \ref{MRD}. With respect to canonical NSs of mass 1.4 M$_{\odot}$, the average density in case-1 increases significantly with the increasing mass, while in the case-3, it slightly decreases with increasing mass.  Again, because the pressure increases with density, to support NSs with the same masses,
in the case where the density is higher the EOS can be softer.

We notice that there are secondary peaks or shoulders in the PDFs of $J_0$ and $J_{\mathrm{sym}}$. This is because of the strong anti-correlation between these two parameters. As discussed in detail in ref. \cite{Xie19}, not only (anti)correlations between parameters of two adjacent terms in either $E_0(\rho)$ or $E_{\rm sym}(\rho)$, cross-(anti)correlations may also exist among parameters used in the two functions. Mathematically one expects to see (anti)correlations between two adjacent terms used to parameterize the same function, e.g, between $K_0$ and $J_0$, or between $L$ and $K_{\rm{sym}}$ when physical conditions are enforced. Physically, for very neutron-rich matter where the isospin asymmetry $\delta$ approaches 1, the $E_0(\rho)$ and $E_{\rm sym}(\rho)\cdot \delta^2$ in Eq. (\ref{eos}) may become equally important in contributing to the total pressure of NS matter. Then, there will be cross-correlations among parameters of $E_0(\rho)$ and $E_{\rm sym}(\rho)$. This most likely happens in dense matter when the symmetry energy becomes very soft (such that the matter is close to pure neutron matter with $\delta=1$).

Shown in Fig. \ref{cor1} are the correlations functions of the high-density EOS parameters for the case-1 (left) and case-3 (right), respectively. In both cases, the anti-correlation between $J_0$ and $J_{\mathrm{sym}}$ is strongest at negative $J_0$ but positive $J_{\mathrm{sym}}$ values. Interestingly, corresponding to the secondary peak/shoulder of their PDFs, there is an appreciable anti-correlation between them at positive $J_0$ but negative $J_{\mathrm{sym}}$ values (where the symmetry energy is super-soft). This is because the required pressure in the high density region to balance the gravity of NSs in the data set can come from either the $J_0$ or $J_{\mathrm{sym}}$ terms. The secondary peak/shoulder in case-1 is stronger than that in case-3 because only in the former case the density is high enough for the $J_{\mathrm{sym}}$ to play an important role and can lead to super-soft symmetry energy. It is also seen that generally two adjacent $E_{\rm sym}(\rho)$ parameters, e.g., $L$ and $K_{\rm{sym}}$,  $K_{\rm{sym}}$ and $J_{\mathrm{sym}}$, are anti-correlated as one expects while the relationships between two distant parameters, e.g.,  $L$ and $J_{\mathrm{sym}}$ or $L$ and $J_{0}$ are more complicated and normally correlated weakly.

\begin{figure*}[htb]
\begin{center}
\resizebox{0.7\textwidth}{!}{
  \includegraphics{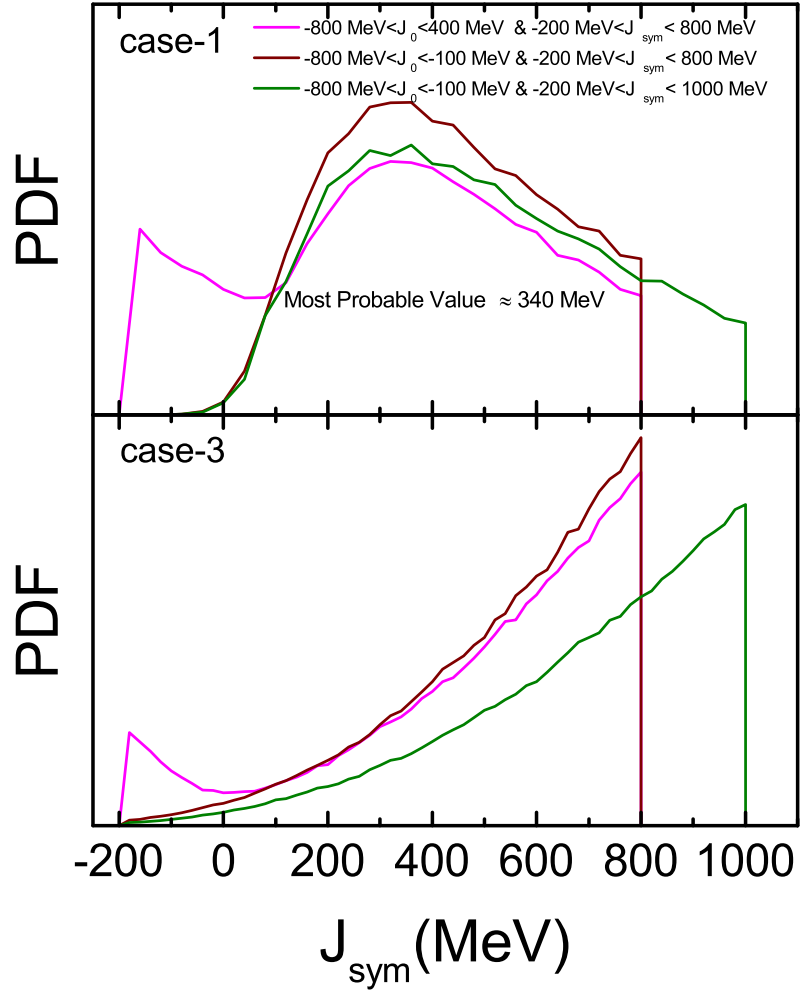}
  }
\caption{Posterior PDFs of $J_{\mathrm{sym}}$ for the case-1 and case-3 by varying the prior limits of $J_0$ and $J_{\mathrm{sym}}$ as indicated. The magenta lines are the results of using the default prior ranges.}\label{jsym13}
\end{center}
\end{figure*}

\subsection{Effects of prior ranges of high-density EOS parameters}
Comparing the PDFs of all EOS parameters in both case-1 and case-3, the most dramatic difference is in the shapes of the PDFs for $J_{\mathrm{sym}}$. For the case-1, it has a major peak indicating that the most probable value of $J_{\mathrm{sym}}$ is around 340 MeV. For the case-3, however, it peaks at the upper boundary at 800 MeV. As we discussed in the introduction, the $J_{\mathrm{sym}}$ parameter characterizing the high-density behavior of nuclear symmetry energy is so far not constrained by any experiment or observation. The prior range we used for it $-200 \leq J_{\rm{sym}}\leq 800$ MeV is completely based on surveys of some theoretical predictions as mentioned before.  While the situation for the high-density SNM matter is better due to the progress in analyzing heavy-ion reaction experiments \citep{Xie20}, the $J_0$ also suffers from large uncertainties. It is thus necessary to examine how these two high-density EOS parameters affect the PDF of the $J_{\mathrm{sym}}$, especially for the case-3 where the peak at the upper boundary of $J_{\mathrm{sym}}=800$ MeV looks suspicuous.

Shown in Fig. \ref{jsym13} are the posterior PDFs of $J_{\mathrm{sym}}$ for the case-1 and case-3 by varying the prior limits of $J_0$ and $J_{\mathrm{sym}}$ as indicated. The magenta lines are the results of using the default prior ranges. By comparing the three calculations in each case, we can see clearly effects of the upper bounds of both
$J_0$ and $J_{\mathrm{sym}}$. Since their PDFs vanish or are very small at their lower boundaries as shown in Fig. \ref{default-l1458} already, it is not necessary to modify the lower boundaries of $J_0$ and $J_{\mathrm{sym}}$. In the case-1, the PDF peak of $J_{\mathrm{sym}}$ remains around $J_{\mathrm{sym}}=340$ MeV, indicating a reliable extract of the most probable value of $J_{\mathrm{sym}}$ although the PDF values vary a little bit at the two ends.
In the case-3, however, the peak or the most probable value of $J_{\mathrm{sym}}$ keeps changing as its upper limit increases, indicating that the data in this case do not constrain the $J_{\mathrm{sym}}$.
This is consistent with our earlier finding that the radius data of canonical NSs do not constrain the high-density symmetry energy above $2\rho_0$ and the corresponding $J_{\mathrm{sym}}$ parameter.
As shown in Fig. \ref{MRD}, the average density reached in massive NSs in case-3 is slightly lower than that reached in canonical NSs. While in the case-1, the average density in NSs of 2.0 M$_{\odot}$ is about
2.3 times that in canonical NSs. Therefore, the data set in case-1 can constrain the $J_{\mathrm{sym}}$ while those in case-3 can't. It is also interesting to see that the second peak near $J_{\mathrm{sym}}=-200$ MeV disappears while the most probable value of $J_{\mathrm{sym}}$ stays around 340 MeV when the $J_0$ is restricted to less than -100 MeV. This is due to the anti-correlation between $J_0$ and $J_{\mathrm{sym}}$ as we discussed in the previous subsection.
\begin{figure*}[htb]
\begin{center}
\resizebox{0.8\textwidth}{!}{
  \includegraphics{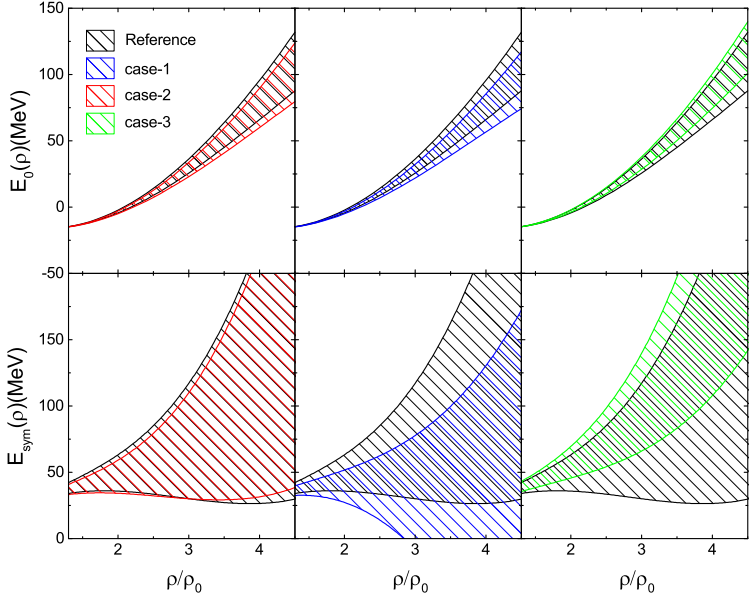}
  }
\caption{Comparisons of the 68\% confidence boundaries of $E_0(\rho)$ and $E_{\mathrm{sym}}(\rho)$ from the four data sets. }\label{EoS-nor}
\end{center}
\end{figure*}
\subsection{EOS confidence boundaries at supra-saturation densities constrained by radii of massive neutron stars}
Applying the obtained posterior PDFs of the EOS parameters in Eqs. \ref{E0para} and \ref{Esympara}, we can easily obtain constraining bands of $E_0(\rho)$ and $E_{\mathrm{sym}}(\rho)$ at any specific confidence level. For example, shown in Fig. \ref{EoS-nor} are the constraining bands on the $E_0(\rho)$ and $E_{\mathrm{sym}}(\rho)$ at 68\% confidence level for the three cases in comparison with the reference from GW170817. The $E_{\mathrm{sym}}(\rho)$ bands for the case-2 and the reference largely overlap, while the $E_0(\rho)$ band for the case-2 is only slightly lower than that of the reference as we expected earlier from examining the PDFs of the EOS parameters. While both the $E_0(\rho)$ and $E_{\mathrm{sym}}(\rho)$ bands for the case-1 are significantly softer than those for the case-3 also as we expected. In particular, due to the large uncertainty of $J_{\mathrm{sym}}$ and the high densities reached in the case-1, the 68\% confidence band for the $E_{\mathrm{sym}}(\rho)$ is very wide in this case. We notice that the lower boundary of the high-density $E_{\mathrm{sym}}(\rho)$ has some dependence on the NS maximum mass used. If 2.14 M$_{\odot}$ instead of 1.97 M$_{\odot}$ is used for the NS maximum mass observed, the lower boundary of $E_{\mathrm{sym}}(\rho)$ around $3\rho_0$ increases slightly \citep{Zhang19epj,Zhang19apj,YZhou19,Xie19}. In the case-3, however, it is seen that the symmetry energy becomes significantly more stiff and is distinctly different from that in the case-1.

Comparing the three cases, it is seen that the constraining bands on their SNM EOSs $E_0(\rho)$ are not much different. This is again because the three cases cover the same range of NS masses determined mainly by the $E_0(\rho)$ with little influence from the $E_{\mathrm{sym}}(\rho)$. Going back to the imaged data sets shown in Fig. \ref{MRD}, this means that a $\pm 15\%$ error bar in measuring the NS mass-radius correlation will not affect the accurate extract of the SNM EOS. On the other hand, the $E_{\mathrm{sym}}(\rho)$ bands for the three cases are rather different, indicating that the main cause for the different mass-radius relations shown in Fig.\ \ref{MRD} is the underlying high-density behavior of nuclear symmetry energy. Thus, a precise measurement of the mass-radius correlation for massive NSs hopefully in the near future will help further constrain the nuclear symmetry energy above $2\rho_0$ with little influence from the remaining uncertainties of the SNM EOS $E_0(\rho)$.

\subsection{Effects of the cubic term in parameterizing the high-density symmetry energy}
In Bayesian kinds of analyses, there is a general question as to how the extracted physical quantities may depend qualitatively and/or quantitatively on the parameterizations used. Ideally, at least the qualitative conclusions should be independent of the parameterizations used. Moreover, correlations among the model parameters may also depend on the total number of parameters used and how well we know about the high-order terms. Of course, considering the limitations of data available and computing costs, since high-order parameters normally involve more uncertainties some compromise may thus have to be made. For example, it was found that some of the existing correlations among different empirical parameters of the nuclear EOS, e.g, between $L$ and $K_{\rm{sym}}$, can be understood from basic physical constraints imposed on the Taylor expansions of $E_0(\rho)$ and $E_{\mathrm{sym}}(\rho)$ at $\rho_0$ \citep{MM3}. However, large dispersions of the correlations among low-order empirical parameters can be induced by the unknown higher-order empirical parameters. For example, the correlation between $E_{\rm sym}(\rho_0)$ and $L$ depends strongly on the poorly known $K_{\rm sym}$, while the correlation between $L$ and $K_{\rm sym}$ is strongly blurred by the even more poorly known $J_0$ and $J_{\rm sym}$ \citep{MM3}. Our results discussed above are in general agreement with these earlier findings. Because the $J_{\rm sym}$ is so poorly known, often it is simply set to zero in many studies in the literature, see, e.g, discussions in several recent works \citep{Margueron18,Baillot19,Perot19,Zimmerman20,Wei20}. To see how our results may depend on the $J_{\mathrm{sym}}$ term,  in the following we
compare the PDFs of EOS parameters and the corresponding confidence boundary of high-density $E_{\mathrm{sym}}(\rho)$ calculated with the default $J_{\mathrm{sym}}$ randomly generated within 200 MeV$\leq J_{\mathrm{sym}} \leq$ 800 MeV and those calculated by setting $J_{\mathrm{sym}}$= 0 MeV.

\begin{figure*}[htb]
\begin{center}
\resizebox{0.8\textwidth}{!}{
  \includegraphics{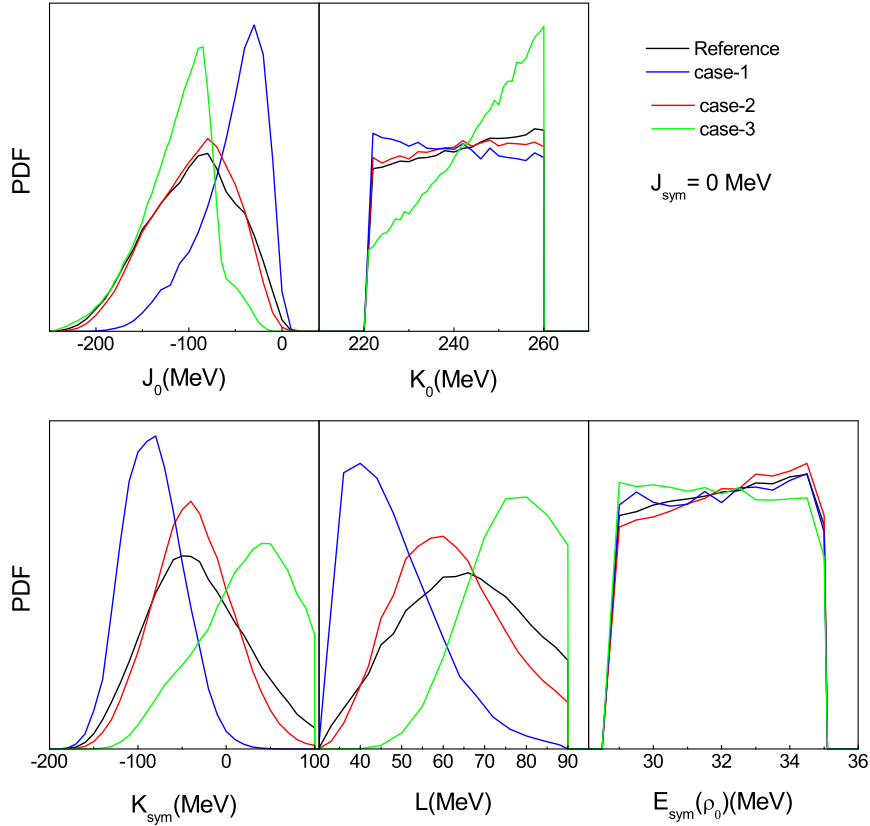}
  }
\caption{Posterior probability distribution functions of EOS parameters from the four sets of mass-dependent radius data of neutron stars by settting $J_{\mathrm{sym}}$ = 0.}\label{default-jsym0}
\end{center}
\end{figure*}

Shown in Fig. \ref{default-jsym0} are the posterior PDFs of the EOS parameters by setting $J_{\mathrm{sym}}$ to zero. Compared to the default results shown in Fig. \ref{default-l1458},
several interesting observations can be made:
\begin{itemize}
\item As for the default results, except the posterior PDF of $K_0$ in the case-3, the PDFs of $K_0$ and $E_{\mathrm{sym}}(\rho_0)$ are less affected by the data sets used.
\item The PDFs of $J_0$, $K_{\mathrm{sym}}$ and $L$ are narrowed down to smaller ranges than the default results when the $J_{\mathrm{sym}}$ has a wide uncertainty range.
\item The differences between results from the case-1 and case-3 become larger
\item The most probable value of $J_0$ shifts significantly to higher values compared to the default results to keep the total pressure the same when the contribution from the high-density symmetry energy is turned off by setting $J_{\mathrm{sym}}$ to zero.
\end{itemize}
\begin{table*}[htbp]
\centering
\caption{Most probable values and their 68\% credible intervals of $J_0$, $K_0$, $K_{\mathrm{sym}}$ and $L$ with 200 MeV$\leq J_{\mathrm{sym}} \leq$ 800 MeV and $J_{\mathrm{sym}}$= 0 MeV, respectively.}\label{MP1}
 \begin{tabular}{lccccccc}
  \hline\hline
  Parameters (MeV) &200 MeV$\leq J_{\mathrm{sym}} \leq$ 800 MeV &$J_{\mathrm{sym}}$= 0 MeV \\
  &Reference, case-1, case-2, case-3 &Reference, case-1, case-2, case-3 \\
  \hline\hline\\
 $J_0:$ &$-165_{-45}^{+55},  -180_{-50}^{+50},  -170_{-40}^{+60}, -100_{-70}^{+20}$ &$-80_{-60}^{+40}, -40_{-30}^{+30}, -80_{-50}^{+40}, -85_{-55}^{+10}$\\
 $K_0:$ &$258_{-24}^{+2},  222_{-0}^{+24},  222_{-0}^{+26}, 260_{-22}^{+0}$ &$258_{-24}^{+2}, 222_{-0}^{+26}, 258_{-24}^{+2}, 260_{-20}^{+0}$   \\
 $K_{\mathrm{sym}}:$ &$-120_{-100}^{+80},  -110_{-120}^{+30}, -100_{-90}^{+70}, -30_{-70}^{+80}$ &$-50_{-40}^{+70}, -80_{-40}^{+20}, -40_{-40}^{+50}, 40_{-50}^{+50}$ \\
 $L:$  &$66_{-20}^{+12}, 38_{-6}^{+18}, 50_{-14}^{+14}, 70_{-16}^{+12}$ &$66_{-15}^{+15}, 40_{-9}^{+10}, 60_{-12}^{+12}, 80_{-12}^{+8}$\\
  \hline
 \end{tabular}
\end{table*}

To be more quantitative in comparing the results, the most probable values and 68\% credible intervals of the EOS parameters are listed in Table \ref{MP1}. Interestingly, the most probable values of $J_0$ and $K_{\mathrm{sym}}$
are most significantly shifted. This is what one expects. As we discussed earlier, the $J_{\mathrm{sym}}$ is most strongly anti-correlated with these two parameters. While the low-density parameter
$L$ is much less directly correlated with the high-density $J_{\mathrm{sym}}$ parameter. As a result, the most probable values of $L$ in all cases studied remain approximately the same in the two calculations.
These findings remind us again that cautions have to be taken in interpreting the EOS parameters inferred from Bayesian analyses using different parameterizations even from the same data set.

\begin{figure*}[htb]
\begin{center}
\resizebox{0.8\textwidth}{!}{
  \includegraphics{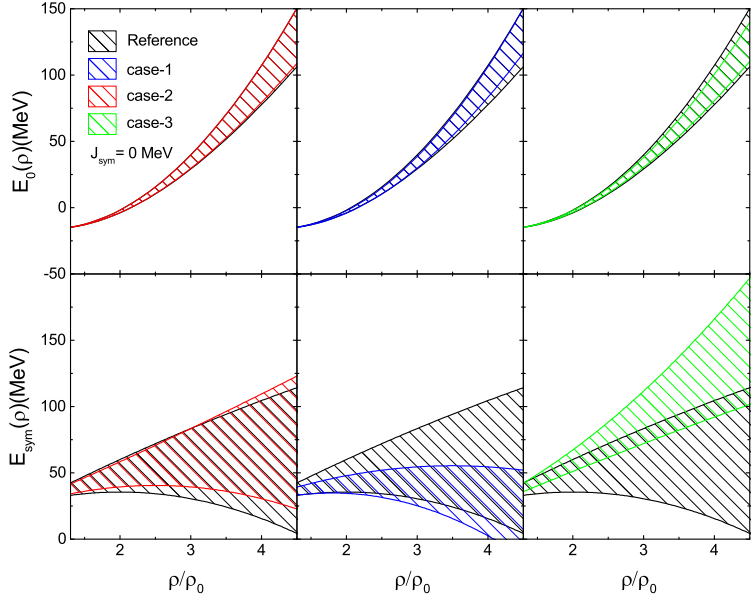}
  }
\caption{The 68\% confidence boundaries of $E_0(\rho)$ and $E_{\mathrm{sym}}(\rho)$ when the parameter $J_{\mathrm{sym}}$ is set to zero. }\label{EoS-jsym0}
\end{center}
\end{figure*}

While the PDFs of some individual EOS parameters depend strongly on how and to what order the EOS is parameterized, the reconstructed posterior EOS from these parameters has less dependence on the parameterization due to the auto-adjustments of the EOS parameters in Bayesian analyses through the likelihood function. Shown in Fig.\ \ref{EoS-jsym0} are the 68\% confidence boundaries of $E_0(\rho)$ and $E_{\mathrm{sym}}(\rho)$ when $J_{\mathrm{sym}}$ is set to zero. Compared to the default results shown in Fig. \ref{EoS-nor}, the SNM EOS in all cases becomes stiffer as already indicated by the increased $J_0$ values discussed above, while the $E_{\mathrm{sym}}(\rho)$ in all cases becomes softer at high densities by setting $J_{\mathrm{sym}}$= 0 MeV. Nevertheless, the relative effects of the different mass-radius correlation in the four data sets remain approximately the same. Namely, the data in the case-1 leads to a significantly softer symmetry energy at high densities than the case-3. Thus, our qualitative conclusion is independent of the EOS parameterizations we used.

\begin{table*}[htbp]
\centering
\caption{Most probable values and the 68\% credible intervals of $E_{\mathrm{sym}}(2\rho_0)$ and $E_{\mathrm{sym}}(3\rho_0)$ with 200 MeV$\leq J_{\mathrm{sym}} \leq$ 800 MeV and $J_{\mathrm{sym}}$= 0 MeV, respectively. }\label{MP2}
 \begin{tabular}{lccccccc}
  \hline\hline
  $E_{\mathrm{sym}}(\rho)$  (MeV) &200 MeV$\leq J_{\mathrm{sym}} \leq$ 800 MeV &$J_{\mathrm{sym}}$= 0 MeV \\
  &Reference, case-1, case-2, case-3 &Reference, case-1, case-2, case-3 \\
  \hline\hline\\
  $E_{\mathrm{sym}}(2\rho_0):$  &$54.8_{-19}^{+8.4}$, $43.2_{-15.9}^{+8.5}$, $50.5_{-16.3}^{+9.4}$, $61.1_{-15.4}^{+8.4}$  &$53.4_{-17.9}^{+6.7}$, $43.4_{-8.9}^{+4.4}$, $52.3_{-12.9}^{+6.1}$, $60.1_{-9}^{+7.2}$\\
 $E_{\mathrm{sym}}(3\rho_0):$  &$91.3_{-61.2}^{+25.8}$, $52.2_{-60.4}^{+25.6}$, $85.1_{-55}^{+24.9}$, $114_{-48}^{+25.8}$  &$66.7_{-36.2}^{+16.7}$, $43.4_{-18.2}^{+11.1}$, $65.6_{-26.2}^{+17.8}$, $92.1_{-20}^{+19.6}$\\\\
 \hline
 \end{tabular}
\end{table*}
To make a more quantitative comparison, listed in Table \ref{MP2} are the most probable values and 68\% credible intervals of $E_{\mathrm{sym}}(2\rho_0)$ and $E_{\mathrm{sym}}(3\rho_0)$ from the two calculations. As we found already in Ref. \citep{Xie19}, the value of $E_{\mathrm{sym}}(2\rho_0)$ is approximately independent of the EOS parameterizations used. This is because around $2\rho_0$ the symmetry energy is mostly controlled by the $L$ and $K_{\rm sym}$ parameters with little influence from the $J_{\mathrm{sym}}$ term mostly through its anti-correlation with $K_{\rm sym}$.  Moreover, the most probable values of $E_{\mathrm{sym}}(2\rho_0)$ from the four different data sets are different from each other by about 20\% but overlap significantly within their $1\sigma$ error bars. Interestingly, they are all consistently with the results shown in Fig. \ref{Esym-survey} within their error bars. For the $E_{\mathrm{sym}}(3\rho_0)$, it is seen that it decreases by about 15\% to 27\% when the $J_{\mathrm{sym}}$ is set to zero. Nevertheless, this is still much smaller than the approximately 53\% difference in $E_{\mathrm{sym}}(3\rho_0)$ between the case-1 and case-3 in both calculations. Therefore, one can still draw qualitatively clear conclusions about the $E_{\mathrm{sym}}(3\rho_0)$ from observed mass-radius correlations of massive NSs regardless of the EOS parameterizations one use in the Bayesian analyses.

Here it is necessary to emphasize that there is no physical reason to ignore the $J_{\mathrm{sym}}$ term besides simplifying calculations. Moreover, as we discussed already, in neutron-rich matter when the
symmetry energy is very soft the isospin asymmetry at $\beta$ equilibrium is close to 1. Then the $J_0$ and $J_{\mathrm{sym}}$ are at the same order and are equally important in contributing to the total pressure in NSs. While our comparisons presented in this subsection are interesting, in our opinion, the default results are more physical and reliable. As to even higher order terms, such as the quartic terms some people included in Taylor expanding nuclear energy density functionals, to our best knowledge, there is so far no meaningful constraints from any experiment/observation and model predictions are even more diverse than for $J_{\mathrm{sym}}$. In our opinion, as long as we stay below about $4\rho_0$ above which the quark-hadron phase transition definitely will happen according to many predictions, parametrizing the EOS up to the cubic term with both $J_0$ and $J_{\mathrm{sym}}$ are sufficient and necessary.

\subsection{Verifying the importance of mass-radius curves in the Bayesian inference of EOS parameters in another way}
To this end, some readers may be still wondering to what extent the results presented above may have nothing to do with the radius constraints but come only from the choice of nuclear matter EOS parameterizations. What would have been the result if we had removed the constraint on the slope of the mass-radius curve? We attempt to address this by carrying out a new calculation using a constant radius $R_{\mathrm{M}}=R_{1.4}=11.9 \pm 3.2$ km at 90\% confidence level for all NSs considered. The mean radius is the same as in the case-2 but with an error bar so large that one can no longer distinguish the three cases shown in Fig. \ \ref{MRD} any more.  We thus essentially removed the constraint on the slope of the mass-radius curve. Namely, at 90\% confidence level the most massive NS of mass 2.0 M$_{\odot}$ considered has the same probability to have a radius of $R_{2.0}=8.7$ km as in the case-1 and $R_{2.0}=15.1$ km as in the case-3, while the most probable radius is kept at $R_{2.0}=11.9$ km as in the case-2. 

\begin{figure*}[htb]
\begin{center}
\resizebox{0.8\textwidth}{!}{
  \includegraphics{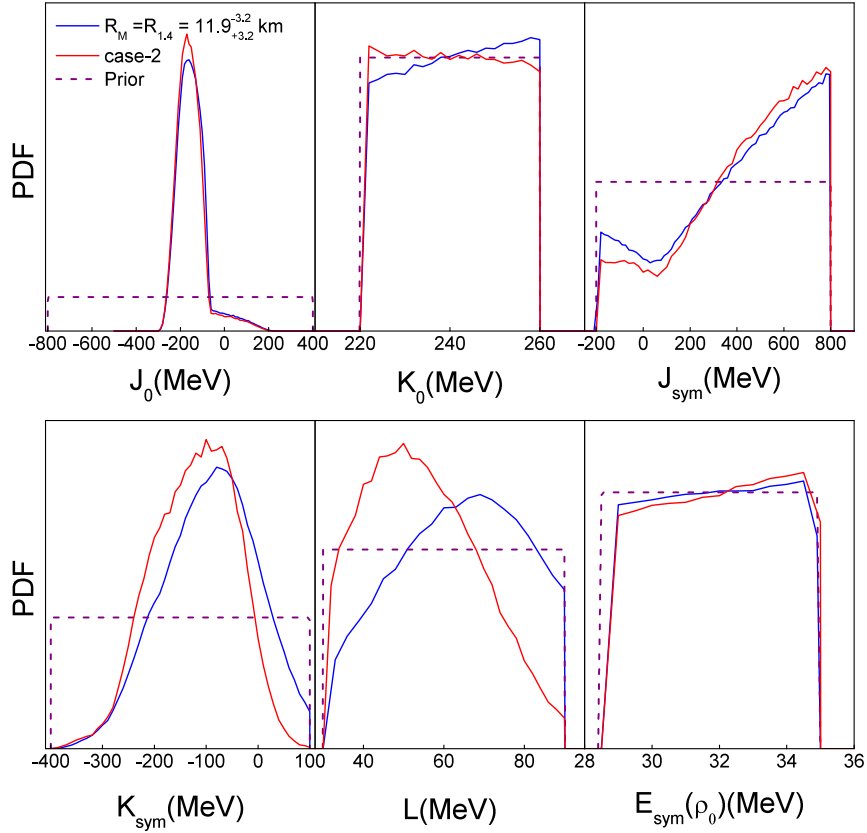}
  }
 \vspace{-0.5cm}
\caption{Prior and posterior probability distribution functions of EOS parameters assuming all neutron stars have the same radius of $R_{\mathrm{M}}=R_{1.4}=11.9 \pm 3.2$ km at 90\% confidence level in comparison with the results from the case-2 (with $R_{\mathrm{M}}=R_{1.4}=11.9 \pm 1.4$ km at 90\% confidence level) shown originally in Fig.4.}\label{newradius}
\end{center}
\end{figure*}

Shown in Fig.\ \ref{newradius} are the prior and posterior PDFs of EOS parameters in the calculation assuming all neutron stars have the same radius of $R_{\mathrm{M}}=R_{1.4}=11.9 \pm 3.2$ km at 90\% confidence level in comparison with the results from the case-2 shown originally in Fig. \ \ref{default-l1458}. First of all, comparing the inferred posterior PDFs with the uniform prior PDFs used as inputs, one can see already very generally the importance of mass-radius measurements in constraining the EOS parameters compared to our prior knowledge. Since this calculation used the same most probable value of $R_{2.0}=11.9$ km as in the case-2 but with a significantly larger error bar (3.2 vs 1.4 km at 90\% confidence level), the differences shown in the results in the two calculations are all due to the difference in the error bars of the radius measurements. The posterior PDFs of all three parameters of the symmetry energy, especially its slope $L$ and curvature $K_{\mathrm{sym}}$, become significantly wider in the new calculation as one expects. On the other hand, as we discussed earlier, the $J_0$ is constrained mostly by the NS maximum mass, its PDF hardly changes in the new calculation as the masses of all NSs considered remain the same. The PDF of $K_0$ shows some slight changes mostly due to its correlations with other parameters. 

Since the radii of all NSs in the two calculations shown in  Fig.\ \ref{newradius} are the same but have two different constant error bars, it is the PDF of the parameter L that is showing the most significant change. This is one might expect from a neutron star radius measurement. However, to answer the questions posted at the beginning of this subsection, one has to compare the new results obtained from using the $R_{\mathrm{M}}=R_{1.4}=11.9 \pm 3.2$ km with those shown earlier from calculations using different mass-radius curves for the case-1 and case-3.
Most interestingly, when we compare the new PDFs of EOS parameters with those in the case-1 and case-3 shown in Fig. 4, it is seen that the PDFs of the symmetry energy parameters in the new calculation are significantly different from those in both the case-1 and case-3, reflecting clearly the importance of the different NS mass-radius curves used in the Bayesian analyses. In particular, the outstanding peak in the PDF of the $J_{\mathrm{sym}}$ parameter in the case-1 where the highest density is reached is completely gone. The new PDF of the $J_{\mathrm{sym}}$ parameter is also significantly different from that in the case-3. We notice that both the $J_0$ and $K_0$ in the new calculation also have appreciably different PDFs compared to those in the case-1 and case-3, again verifying the importance of the mass-radius curves of massive NSs in the Bayesian inference of the dense neutron-rich matter EOS. We are thus very confident that the posterior PDFs of EOS parameters presented above reflect faithfully the NS radius constraints and they are not simply from our choice of nuclear matter EOS parameterizations.

\section{Summary and outlook}
In summary, using an explicitly isospin-dependent parametric EOS of nucleonic matter within the $npe\mu$ mode for the core of NSs, we performed Bayesian analyses using three different sets of imagined mass-radius correlation data of massive NSs. Using the PDFs of EOS parameters as well as the corresponding symmetry energy $E_{\mathrm{sym}}(\rho)$ and SNM EOS $E_0(\rho)$ inferred from GW170817 and NICER radius data for canonical NSs as references, we investigated how future measurements of massive NS radii will improve our current knowledge about the EOS of super-dense neutron-rich nuclear matter. The three imagined radius data sets represent typical predictions using EOSs from various nuclear many-body theories, i.e, the radius stays the same, decreases or increases with increasing NS mass within $\pm 15\%$ between 1.4 M$_{\odot}$ and 2.0 M$_{\odot}$. These three cases model three possible scenarios in the core of NSs assuming no hadron-quark phase transition and/or new particle production: the average density increases quickly, slowly or slightly decreases as the mass increases from 1.4 M$_{\odot}$ to 2.0 M$_{\odot}$. In these three cases the high-density symmetry energy plays different roles. Consequently, the PDFs of EOS parameters and the corresponding EOS confidence boundaries inferred from the three radius data sets are rather different. In particular, while the SNM EOS $E_0(\rho)$ inferred from three data sets are approximately the same, the corresponding high-density symmetry energies $E_{\mathrm{sym}}(\rho)$ at densities above about $2\rho_0$ are very different, indicating that the radii of massive NSs carry important information about the high-density behavior of nuclear symmetry energy with little influence from the remaining uncertainties of the SNM EOS $E_0(\rho)$.
We have also investigated correlations among the EOS parameters and effects of turning on/off the high-order term in parameterizing the symmetry energy. We found that it is important to keep the cubic term to extract more accurately the symmetry energy below about $4\rho_0$.

The major shortcoming of this work is that the NS model used is the minimum $npe\mu$ model without considering phase transitions as well as productions of hyperons and/or baryon resonances that are expected to appear above certain high densities. Nevertheless, as evidenced by many earlier and recent publications in the literature, researches within the minimum NS model provide a useful guidance for possible advanced studies. Extending the present work by incorporating the hadron-quark phase transition and more particles is on our working plan.

Besides the ongoing NICER mission measuring simultaneously the radii and masses of several NSs as well as various gravitational wave searches which can potentially reveal both the masses and radii of super/hyper-massive remnants of NS mergers from multimessengers released, new ideas have been put forward in the Astro 2020 Decadal Survey to measure more accurately the radii of massive NSs using the next-generation X-ray observatories \citep{wp1,Strobe,wp2,Watts19}. It is thus very hopeful that precise mass-radius data for more massive NSs will be available in the near future. On the other hand, new radioactive beam facilities being built around the world \citep{NAP2012,LRP2015,NuPECC} provide great opportunities to probe the EOS of super-dense neutron-rich nuclear matter in controlled laboratory conditions.
Ongoing efforts in nuclear physics, see, e.g., refs. \citep{FRIB,Wolfgang}, are proving complementary information about the EOS of super-dense neutron-rich nuclear matter. Eventually, a truely multimensenger approach involving astrophysics observation, nuclear physics experiments and related theories will enable us to finally pin down the EOS, especially the symmetry energy, of super-dense neutron-rich nuclear matter.\\

\noindent{\bf Acknowledgments:} We thank Bao-Jun Cai, Lie-Wen Chen and Nai-Bo Zhang for very helpful discussions. This work is supported in part by the U.S. Department of Energy, Office of Science, under Award Number DE-SC0013702, the CUSTIPEN (China-U.S. Theory Institute for Physics with Exotic Nuclei) under the US Department of Energy Grant No. DE-SC0009971.

\end{document}